\documentclass[11pt]{article}

\usepackage[final]{acl}
\usepackage{rotating} 
\usepackage{float}
\usepackage[table]{xcolor}
\usepackage{adjustbox}
\newcommand{\ninept}{\fontsize{9pt}{11pt}\selectfont}

\usepackage{amssymb}
\usepackage{pifont}

\usepackage{times}
\usepackage{latexsym}
\usepackage{booktabs,tabularx,multirow}
\usepackage[T1]{fontenc}
\usepackage[ruled,vlined]{algorithm2e} 
\usepackage{amsmath}
\usepackage[utf8]{inputenc}

\usepackage{microtype}

\usepackage{inconsolata}

\usepackage{graphicx}
\usepackage{tikz}

\title{FSA-GRPO: Teaching Auditory LLMs to Use Few-Shot Demonstrations}

\author{
 \textbf{Haolong Zheng\textsuperscript{1}} ~
 \textbf{Siyin Wang\textsuperscript{2}} ~
 \textbf{Xulin Fan\textsuperscript{1}} ~
 \textbf{Zengrui Jin\textsuperscript{2}} ~
 \textbf{Mark Hasegawa-Johnson\textsuperscript{1}} ~
\\
 \textsuperscript{1}University of Illinois Urbana-Champaign, Urbana, IL, USA \\
 \textsuperscript{2}Tsinghua University, Beijing, China
\\ 
 \small{
    haolong2@illinois.edu, jhasegaw@illinois.edu
 }
}
\usepackage{cleveref}

\begin{document}
\maketitle
\begin{abstract}

Few-shot prompting provides an effective way to adapt auditory large language models to low-resource tasks such as children’s speech recognition. However, most auditory large language models are not explicitly trained to perform inference in this demonstration-conditioned format, limiting the extent to which they can benefit from In-Context Learning (ICL). To address this limitation, we introduce Few-Shot Aware GRPO (FSA-GRPO), an RL-based post-training recipe that uses a specially designed reward to encourage the model to leverage few-shot demonstrations, thereby strengthening its few-shot adaptation ability. Notably, training with only 2k high-resource adult ASR utterances improves the model’s general few-shot adaptation ability, yielding gains not only in children’s speech recognition (53.9\% relative WER reduction without any in-domain training) but also in multilingual ASR (including low-resource languages), speech translation, and audio understanding. We further study data selection and the weight and similarity cutoffs of the auxiliary reward to identify an effective training recipe. Our experiments show that when in-domain data are unavailable or cannot be used for training, FSA-GRPO is more effective than direct tuning on related out-of-domain data.

\end{abstract}

\begin{figure*}[ht]
    \centering
    \includegraphics[width=1\linewidth]{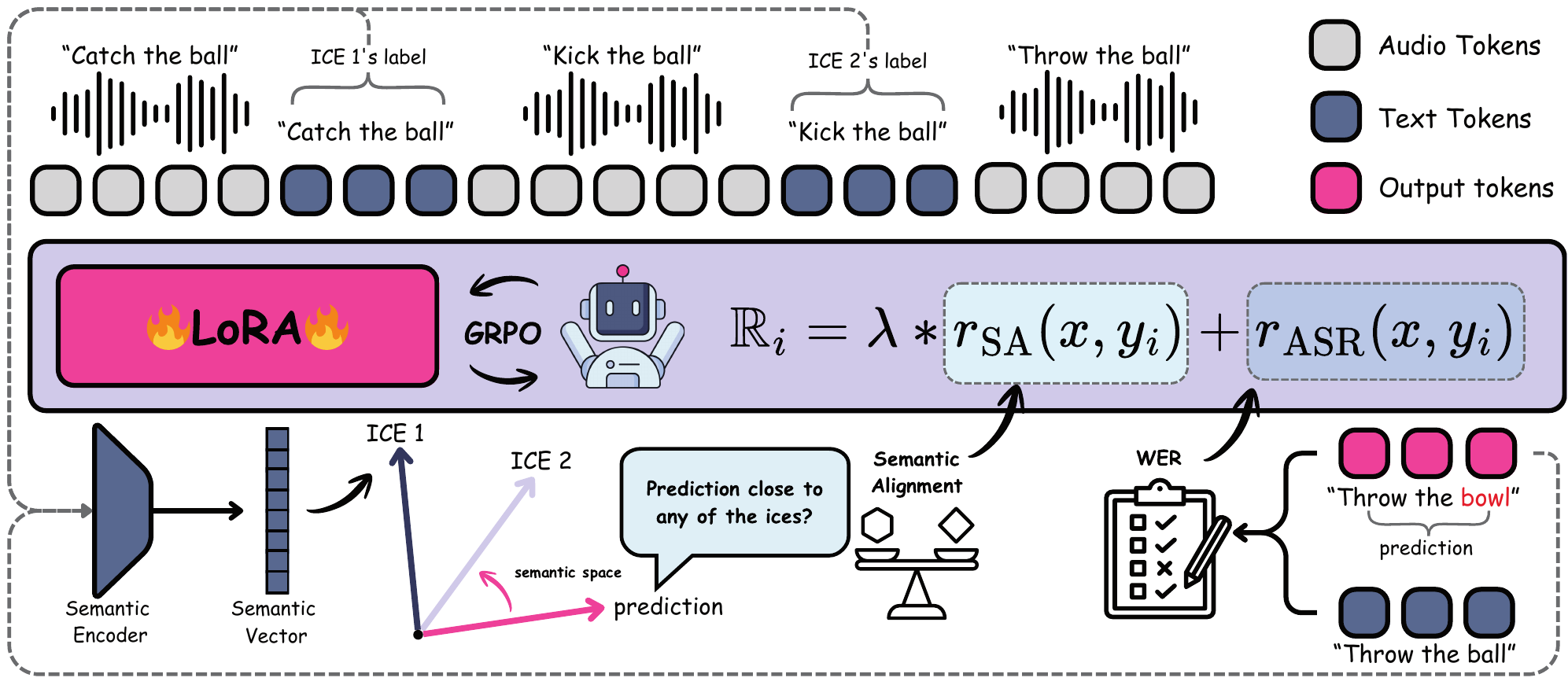}
    \caption{
Overview of \textbf{FSA-GRPO}. 
We construct few-shot ASR training instances by pairing a query utterance with retrieved speech--text in-context examples (ICEs), then optimize a LoRA-augmented auditory LLM with GRPO.
Each sampled response receives a reward combining ASR accuracy, measured by WER against the reference transcript, and semantic alignment, which encourages the prediction to remain close to relevant ICE labels in embedding space. 
By explicitly rewarding both transcription correctness and demonstration awareness, FSA-GRPO teaches the model to better leverage few-shot examples for low-resource and unseen speech/audio tasks.
}
    \label{fig:overview}
\end{figure*}

\section{Introduction}
Recent progress in auditory large language models (LLMs) has led to rapid improvements across a wide range of speech and audio tasks, including automatic speech recognition, speech translation, and audio understanding/reasoning~\cite{Qwen2.5-Omni,ghosh2026audio,wang2026end,wang2026towards}. However, these advances are largely driven by training on large-scale, high-resource data. Their performance can drop substantially in low-resource settings where labeled in-domain data are scarce, expensive to collect, or difficult to release due to privacy and access constraints~\cite{fong25_interspeech}. In such cases, domain-adaptive post-training is often needed to achieve acceptable performance. In practice, the collected data may fail to capture the full variability of the target distribution, leaving a persistent performance gap.
For example, children's speech differs substantially from adults' in acoustic and linguistic patterns, making adult-trained speech models difficult to apply directly to child speech~\cite{ghai2010exploring,gerosa2009review,shivakumar2020transfer}. Although several domain adaptation techniques have been explored under limited-data conditions, child ASR still lags behind standard adult ASR~\cite{fan2024benchmarking}, highlighting the difficulty of adapting speech models when the available target-domain data are sparse and under-representative.

In-Context Learning (ICL)~\cite{brown2020language} allows LLMs to adapt to out-of-domain data by conditioning on a small set of labeled in-domain examples, without requiring gradient updates. Compared with direct fine-tuning on a small amount of low-resource data, which can be brittle under domain shift, ICL allows the model to dynamically adapt to new domains by simply changing the few-shot demonstrations provided at inference time. This few-shot adaptation is particularly useful in low-resource scenarios. However, the gains from vanilla ICL are intuitively upper-bounded because most auditory LLMs are not explicitly exposed to this demonstration-conditioned format during training. A natural solution~\cite{zheng2026metasiclglobalizingauditoryllms} is to post-train auditory LLMs on few-shot formatted data so that they become more familiar with demonstration-conditioned inference. However, this strategy introduces two risks: the model may sacrifice its zero-shot capability after additional tuning, or it may only learn the structure of the few-shot format for the training task rather than learning to use the demonstrations to guide the query prediction in general. In order to benefit more effectively from few-shot examples in low-resource scenarios, we ask an important question:


\textit{Can few-shot adaptation be optimized without catastrophic forgetting of zero-shot inference?}

To answer this question, we propose \textbf{Few-Shot Aware GRPO (FSA-GRPO)}, an RL-based post-training strategy designed to strengthen the model's ICL capability while maintaining direct inference performance. Specifically, we construct few-shot training instances from high-resource ASR data and introduce an auxiliary reward that encourages the model to leverage these demonstrations during generation in addition to the base reward. Our key contributions are summarized as follows:

\begin{itemize}
    \item We propose the first RL-based post-training approach for improving speech in-context learning. The recipe is architecture-agnostic in that it requires only a demonstration-conditioned prompt format and a rule-based reward, and we demonstrate it on two auditory LLMs; broader cross-model effectiveness remains to be established. Training on English ASR data alone improves the model's general ability to exploit demonstrations, and this transfers to held-out tasks and unseen languages: multilingual ASR across 57 FLEURS languages, speech translation in seven CoVoST2 directions, and audio understanding.
    \item Under a rigorous low-resource ASR evaluation setting, we show that FSA-GRPO provides a more effective adaptation strategy than directly fine-tuning the model on related out-of-domain ASR data.
    \item We study two practical design dimensions of FSA-GRPO: selecting high-quality few-shot training instances to improve training efficiency, and analyzing the auxiliary demonstration-aware reward through controlled ablations of its weight and similarity cutoffs.
\end{itemize}


\section{Related Work}

\paragraph{Few-Shot Adaptation for Speech and Audio.}
ICL-based few-shot adaptation has been explored beyond text and has shown promise across multimodal settings, including vision-language and audio-language models~\cite{huang2023language, kong2024audio, coreteam2025mimoaudio}. In the speech domain, prior work has mainly studied ICL as an inference-time adaptation mechanism. For ASR, the first SICL work adapted Whisper to Chinese dialect recognition by conditioning on a few paired speech utterances and transcripts from the target dialect or speaker, showing gains in dialect-level, speaker-level, and spontaneous continuous-speech settings~\cite{wang2024can}. Subsequent work has mainly focused on selecting more useful demonstrations: Bayesian example selection was evaluated on Chinese dialect ASR and regional English ASR~\cite{wang2024bayesian}, while M2R-Whisper combines sentence-level ICL with token-level retrieval to improve Mandarin and Mandarin-subdialect ASR in low-resource settings~\cite{zhou2025m2r}. More recent LMM-based SICL methods extend this paradigm beyond Whisper: TICL retrieves semantically related demonstrations for accented English, multilingual, and children's ASR~\cite{zheng2025ticltextembeddingknnspeech}, and TICL+ further adds acoustic reranking for children's speech recognition~\cite{zheng2025ticlcasestudyspeech}. Beyond ASR, demonstration-conditioned speech/audio models have also been studied for speech translation, speech classification, and emotion-related adaptation. COSMIC and SALM use speech-text demonstrations or speech-supervised in-context training for speech-to-text translation and keyword-aware ASR/AST adaptation~\cite{pan2023cosmic, chen2024salm}. In speech classification, textless speech LMs have been warmup-trained to solve unseen classification tasks from utterance-label demonstrations~\cite{chang2024exploring}, and recent SLU work studies in-context learning for unseen spoken-language-understanding tasks without task-specific annotations~\cite{agrawal2025spoken}. For affective speech, Ihori et al. personalize speech emotion recognition by conditioning on a few emotional utterances from the target speaker~\cite{ihori2025few}, while UniAudio 1.5 further extends cross-modal ICL to broader audio tasks including speech emotion classification, audio classification, TTS, and speech enhancement~\cite{yang2024uniaudio}. Together, these studies suggest that few-shot prompting is a flexible and effective adaptation mechanism for speech and audio tasks, but most of them focus on inference-time prompting or retrieval rather than explicitly optimizing the model's ability to use demonstrations.

\paragraph{Training Models to Use Demonstrations.}
A complementary line of work studies whether the ability to learn from demonstrations can itself be improved through training. MetaICL~\cite{min2022metaicl} enhances LLMs' ICL ability through few-shot meta-training on diverse NLP tasks. However, because it operates entirely in the text modality, it remains unclear whether this strategy can transfer to speech-conditioned LLMs.
SMILE~\cite{hsu2024smile} and Omnilingual ASR~\cite{omnilingual2025omnilingual} further explore ICL-style training in the speech domain. SMILE uses high-resource ASR data to perform ICL-style tuning on Whisper, while Omnilingual ASR explicitly trains the language decoder in a few-shot format for multilingual ASR, enabling transfer to unseen languages by conditioning on a few examples. However, both methods are tied to specific ASR-oriented model designs, leaving unclear whether a similar paradigm can be transferred to general auditory LLMs. \citet{li2026multimodal} further extends this direction to speech LLMs. Their cross-lingual setting shows that ICL-format training can generalize to unseen low-resource languages, but it remains unclear whether it can transfer to broader audio tasks. Most related to our work, MetaSICL~\cite{zheng2026metasiclglobalizingauditoryllms} studies whether ICL-style post-training with various speech tasks can improve few-shot adaptation for auditory LLMs. It shows that meta-training with demonstration-conditioned speech tasks can improve performance on held-out tasks. However, its analysis primarily focuses on few-shot performance, leaving unclear whether such training improves ICL behavior without sacrificing the model's direct inference ability.


\paragraph{RL Post-Training for Speech and Audio LLMs.}
Group Relative Policy Optimization (GRPO) was introduced in DeepSeekMath as a memory-efficient variant of PPO~\cite{schulman2017proximal,shao2024deepseekmath} and later became a common framework for rule-based RL post-training of LLMs~\cite{guo2025deepseekr1}. Unlike supervised fine-tuning, GRPO can directly optimize sequence-level, non-differentiable objectives such as exact match, BLEU, or WER through automatic rewards. Recent speech and audio studies have applied this paradigm to audio question answering, spoken question answering, speech translation, and ASR, often showing stronger performance than SFT under limited post-training data or improved robustness under domain shift~\cite{li2025reinforcement,elmakies2025advancing,shivakumar2025grpospeech}.

\section{Methodology}

We formulate FSA-GRPO as a demonstration-conditioned RL post-training recipe. Given a query audio and a set of retrieved (audio, text) demonstrations, the model is optimized to generate the query transcript while receiving rewards for both transcription correctness and demonstration semantic alignment. Figure~\ref{fig:overview} illustrates the overall framework, and Algorithm~\ref{alg:training-overview} summarizes the training procedure.

\subsection{Training Data \& Preparation}
\label{sec:training-data}

We construct training data using only the English subset of Common Voice~\cite{commonvoice:2020}. This design highlights a key advantage of our recipe: it does not require scarce or privacy-sensitive target-domain data for training, but instead learns few-shot adaptation behavior from high-resource ASR data. Since GRPO training is computationally expensive, we sample 2k query instances using the data-selection strategy identified in Section~\ref{sec:data-selection}, unless otherwise specified.


To mirror the few-shot inference format, we construct training instances following TICL~\cite{zheng2025ticltextembeddingknnspeech}. Specifically, we treat the Common Voice development split as the query set $\mathcal{D}_{\text{query}}$ and training split as the demonstration pool $\mathcal{D}_{\text{pool}}$. At each training step, we sample a query utterance-transcript pair $(x_q, y_q) \sim \mathcal{D}_{\text{query}}$ and retrieve $k$\footnote{We use $k=3$ throughout the paper.} in-context demonstrations ${(x_j, y_j)}_{j=1}^{k}$ from $\mathcal{D}_{\text{pool}}$. The model is trained to generate the query transcript conditioned on the retrieved demonstrations and the query audio, i.e., $P_{\theta}(y_q \mid x_1, y_1, \ldots, x_k, y_k, x_q)$. This exposes the model to the same demonstration-conditioned format used at test time.

\subsection{Reward Design}

For GRPO training, we use rule-based rewards derived from ASR supervision and the retrieved in-context demonstrations. Given a prompt, the policy samples a group of $G$ candidate responses $g(x)=\{y_i\}_{i=1}^{G}$. Following GRPO, each response is assigned a reward $\mathbb{R}_i$, and the corresponding advantage is computed relative to the reward distribution within the same group:
\begin{equation}
\hat{A}_i =
\frac{\mathbb{R}_i - \frac{1}{G}\sum_{j=1}^{G}\mathbb{R}_j}
{\sigma\left(\{\mathbb{R}_j\}_{j=1}^{G}\right)}.
\label{eq:grpo_advantage}
\end{equation}

\paragraph{ASR Reward.}
The primary reward measures transcription accuracy using word error rate (WER):
\begin{equation}
r_{\mathrm{ASR}}(x,y_i)
=
\max\left(
0,\,
1-\mathrm{WER}(\tilde{y}_i,\tilde{y}^{\star})
\right),
\label{eq:asr_reward}
\end{equation}
where $\tilde{y}_i=\mathcal{N}(y_i)$ and $\tilde{y}^{\star}=\mathcal{N}(y^\star)$ denote the normalized prediction and reference transcript.
The ASR reward allows the model to take advantage of in-context examples when they are useful and ignore them when they are not, thereby helping optimize few-shot adaptation without catastrophic forgetting of zero-shot inference.

\paragraph{Semantic Alignment Reward.}
Prior work has shown that auditory LLMs benefit more from semantically similar in-context examples than from arbitrary demonstrations~\cite{zheng2025ticltextembeddingknnspeech}. Motivated by this observation, we introduce a semantic alignment (SA) reward to encourage the model to generate responses that remain semantically aligned with the retrieved demonstrations.

Let $\mathcal{I}(x)$ denote the \emph{index set} of the $k$ demonstrations retrieved for input $x$, whose transcripts are $\{y_j\}_{j\in\mathcal{I}(x)}$. To keep generations and demonstrations distinct, we write $\hat{y}_i$ for the $i$-th sampled response and $y_j$ for a retrieved demonstration transcript. We encode both with the same sentence encoder $f_\phi$~\cite{reimers-2019-sentence-bert} and normalize their embeddings:
\[
\begin{aligned}
z_i &= \frac{f_\phi(\hat{y}_i)}
            {\lVert f_\phi(\hat{y}_i)\rVert_2},\\
e_j &= \frac{f_\phi(y_j)}
            {\lVert f_\phi(y_j)\rVert_2},
\qquad j \in \mathcal{I}(x).
\end{aligned}
\]
Not all retrieved demonstrations are equally informative for a given query, so we compute the SA reward using only the closest demonstration in the embedding space. The semantic similarity between the generated response and the retrieved demonstrations is therefore computed as
\[
s_i =
\max_{j\in\mathcal{I}(x)}
z_i^\top e_j .
\]

Since the SA reward is intended as a lightweight auxiliary signal rather than the primary optimization target, we bound its value with thresholds to encourage demonstration awareness while avoiding over-copying from the retrieved labels. We convert this similarity into a bounded reward using a linear ramp:
\begin{equation}
r_{\mathrm{SA}}(x,y_i)
=
\operatorname{clip}
\left(
\frac{
s_i-\tau_{\mathrm{low}}^{g(x)}
}{
\tau_{\mathrm{full}}^{g(x)}-\tau_{\mathrm{low}}^{g(x)}
},
0,1
\right),
\label{eq:sa_reward}
\end{equation}
where $\tau_{\mathrm{low}}^{g(x)}$ and $\tau_{\mathrm{full}}^{g(x)}$ are group-specific thresholds. Similarities below $\tau_{\mathrm{low}}^{g(x)}$ receive zero SA reward, while similarities above $\tau_{\mathrm{full}}^{g(x)}$ receive full SA reward. 
By default, we set $\tau_{\mathrm{low}}^{g(x)}$ and $\tau_{\mathrm{full}}^{g(x)}$ to the 25th and 50th percentiles, respectively, of the group-specific empirical similarity distribution; Section~\ref{sec:weight-ablation} analyzes alternative cutoff settings.

The final reward combines transcription accuracy and demonstration relevance:
\begin{equation}
 \mathbb{R}_i = \lambda * r_{\mathrm{SA}}(x,y_i) + r_{\mathrm{ASR}}(x,y_i),
\label{eq:final_reward}
\end{equation}
where $\lambda$ controls the weight of the SA reward.

\subsection{GRPO Details \& Model Selection}


We apply the proposed GRPO-based post-training recipe to both $\textit{Qwen2.5-Omni}$~\cite{Qwen2.5-Omni} and $\textit{Audio-Flamingo-Next}$~\cite{ghosh2026audio}. 
To keep training lightweight and reduce overfitting, we freeze the audio encoder tower and update only LoRA~\cite{hu2022lora} adapters inserted into all linear layers of the language backbone. 
Unless otherwise specified, we use LoRA rank 8 and alpha 32.

For GRPO rollouts, we sample 8 candidate responses per prompt, corresponding to group size $G=8$, with $\text{temperature}=1.0$, $\text{top-}p=1.0$, and no top-$k$ filtering. 
The KL reference policy $\pi_{\mathrm{ref}}$ is the frozen base model with LoRA adapters disabled; no separate reference model is instantiated. 
We use KL coefficient $\beta=0.001$, clipping parameter $\epsilon=0.2$, learning rate $5\times10^{-5}$, and the \texttt{adamw\_torch\_fused} optimizer with betas $(0.9, 0.95)$, epsilon $10^{-8}$, and weight decay $0.1$. 
Training uses a cosine scheduler without warmup. 
The final model is selected as the checkpoint with the best training reward.
All GRPO training is conducted on NVIDIA A100 40GB GPUs.

\begin{figure*}[t]
    \centering
    \includegraphics[width=1\linewidth]{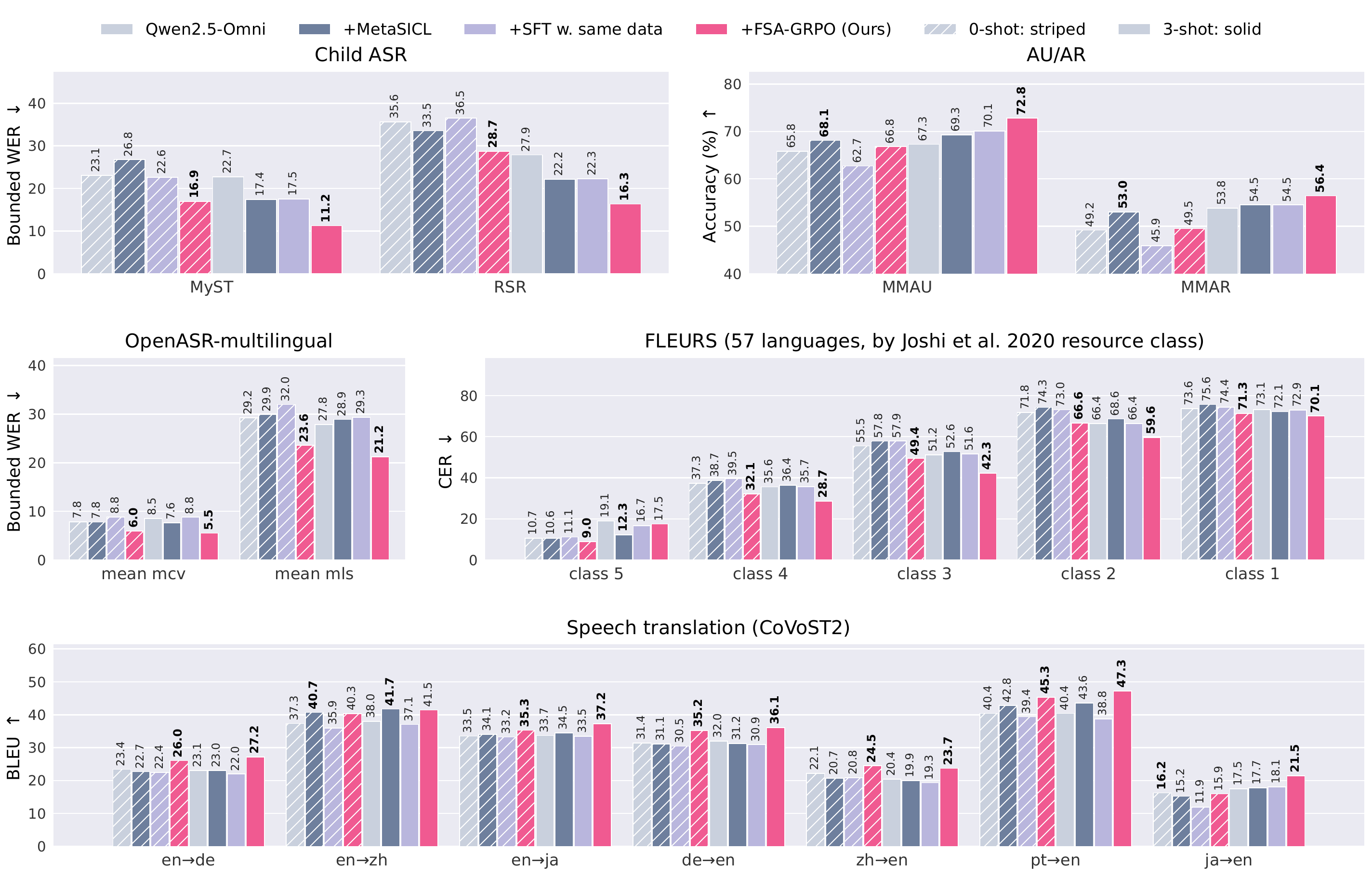}
    \caption{
Main results for $\textit{Qwen2.5-Omni}$, MetaSICL, direct SFT on the same training data, and FSA-GRPO.
Striped bars are zero-shot inference; solid bars are three-shot inference with in-context examples.
Lower is better for WER and CER, higher for accuracy and BLEU; FLEURS languages are grouped by \citet{joshi-etal-2020-state} resource class (5 = highest).
Bold marks the best score among the four systems, within each shot setting separately.
FSA-GRPO improves few-shot performance in nearly every setting, transferring beyond child ASR to multilingual recognition and translation despite English-only training.
The one cell where the base model leads, ja$\rightarrow$en zero-shot, reverses under COMET~\citep{rei-etal-2020-comet,rei-etal-2022-comet} (78.5 vs.\ 76.6).
Full per-language and per-direction results are in Appendix~\ref{appendix:full-results}.
}
    \label{fig:baseline}
\end{figure*}

\subsection{Evaluation Data \& Metrics}

Following MetaSICL~\cite{zheng2026metasiclglobalizingauditoryllms}, we use child ASR (MyST~\cite{pradhan2024my}, RSR~\cite{ai4exceptionaled_rsr_hf}) as the main low-resource task, and hold out audio understanding/reasoning (MMAU~\cite{sakshi2025mmau}, MMAR~\cite{ma2026mmar}), multilingual ASR (Common Voice~\cite{commonvoice:2020}), and speech translation (CoVoST2~\cite{wang2020covost}) to test whether the learned few-shot behavior generalizes beyond the training setup. Beyond that protocol, we additionally evaluate on the Common Voice and MLS~\cite{pratap2020mls} subsets of the Open ASR multilingual leaderboard~\cite{srivastav2025openasr}, on all 57 FLEURS~\cite{conneau2023fleurs} languages for which we hold a retrieval pool, and on five further CoVoST2 directions, which broadens language coverage from three languages and one translation pair to a substantially wider set. We report WER for alphabetic ASR, CER for Chinese and for FLEURS, accuracy for MMAU and MMAR, and BLEU~\cite{papineni-etal-2002-bleu} with up to 4-gram precision for translation, computed with SacreBLEU~\cite{post-2018-call}, applying the Whisper text normalizer~\cite{radford2023robust} to predictions and references before scoring. For child ASR we report bounded WER, in which the per-utterance edit count is capped at the reference length so that severe hallucinations cannot dominate the corpus-level score~\cite{zheng2025interspeech}. Appendix~\ref{appendix:eval-data} describes each evaluation set in full.

\section{Experiments}






\providecommand{\ninept}{\fontsize{9pt}{11pt}\selectfont}
\newcolumntype{Y}{>{\centering\arraybackslash}X} 

We organize the experiments into five parts. First, we compare FSA-GRPO with the main baselines under both zero-shot and few-shot inference settings. Second, we examine a practical low-resource setting, comparing our method with direct-tuning strategies and testing how far the few-shot gain survives when the demonstrations stop matching the query. Third, we ask whether the recipe is specific to one backbone by applying it to a second auditory LLM. Fourth, given the high cost of GRPO training, we study how to construct effective training subsets under a limited budget. Finally, we ablate the two hyperparameters of the auxiliary semantic-alignment reward, its weight and its cutoff.

\subsection{Comparison with Baselines}

\paragraph{Baseline Setup.}
We compare FSA-GRPO with three main baselines. The first baseline is the original $\textit{Qwen2.5-Omni}$ model without post-training, which measures the model's native zero-shot and few-shot capability. The second baseline is MetaSICL~\cite{zheng2026metasiclglobalizingauditoryllms}, a meta-learning-style speech in-context learning method that trains the model with the same few-shot format used at inference time. MetaSICL uses supervised fine-tuning over multiple speech and audio tasks, providing a strong baseline for testing whether meta-style few-shot training alone can improve speech in-context learning. The third baseline is direct SFT with a budget of few-shot ASR training data comparable to that used for FSA-GRPO. This baseline uses the same demonstration-conditioned training format, but replaces our GRPO objective with standard supervised fine-tuning, serving as a direct comparison between SFT-based and RL-based post-training under a comparable data budget.


\paragraph{Overall Results.}

As shown in Figure~\ref{fig:baseline}, FSA-GRPO achieves the strongest overall few-shot performance across the evaluated tasks, outperforming the original model, MetaSICL, and direct SFT on most settings. Importantly, the improvement also transfers to non-overlapping held-out tasks, including audio understanding/reasoning, multilingual ASR, and speech translation, suggesting that FSA-GRPO improves the model's general in-context adaptation ability rather than only fitting the training task. At the same time, FSA-GRPO better preserves zero-shot performance compared with other post-training baselines, indicating that FSA-GRPO strengthens few-shot behavior without making the model overly dependent on in-context examples.

\subsection{Comparison with Direct Tuning}

\paragraph{Motivation \& Experimental Design.}

We evaluate FSA-GRPO in a realistic low-resource ASR scenario where target-domain training data are unavailable. Using RSR child speech as the target domain, we ask a practical question: \textit{when RSR training data cannot be used, should we use direct ASR post-training or train the model with our recipe for few-shot inference?}

We include both related-domain and out-of-domain direct-tuning baselines\footnote{All direct-tuning baselines are trained with standard single-turn ASR prompts.}, which represent conventional direct ASR tuning, while our method explicitly trains the model in a demonstration-conditioned format. For related-domain tuning, we use only 2k MyST\footnote{MyST is also child speech, so it provides a related-domain but still non-target-domain adaptation source.} training examples for direct SFT and direct GRPO.  For out-of-domain data, we use Common Voice, a generic ASR dataset, under two budgets: CV 2k and CV 20k. This design separates three intuitive alternatives: low-resource related-domain direct tuning, low-resource out-of-domain direct tuning, and higher-resource out-of-domain direct tuning.

At inference time, we evaluate each system in two settings. The first uses RSR samples as in-context examples; the second is a stricter out-of-domain setting, where the in-context examples are drawn from MyST. This setting removes all dependence on RSR data and tests whether FSA-GRPO remains useful when no target-domain data is available for either training or demonstrations.



\begin{figure*}[t]
\centering
\includegraphics[width=1\linewidth]{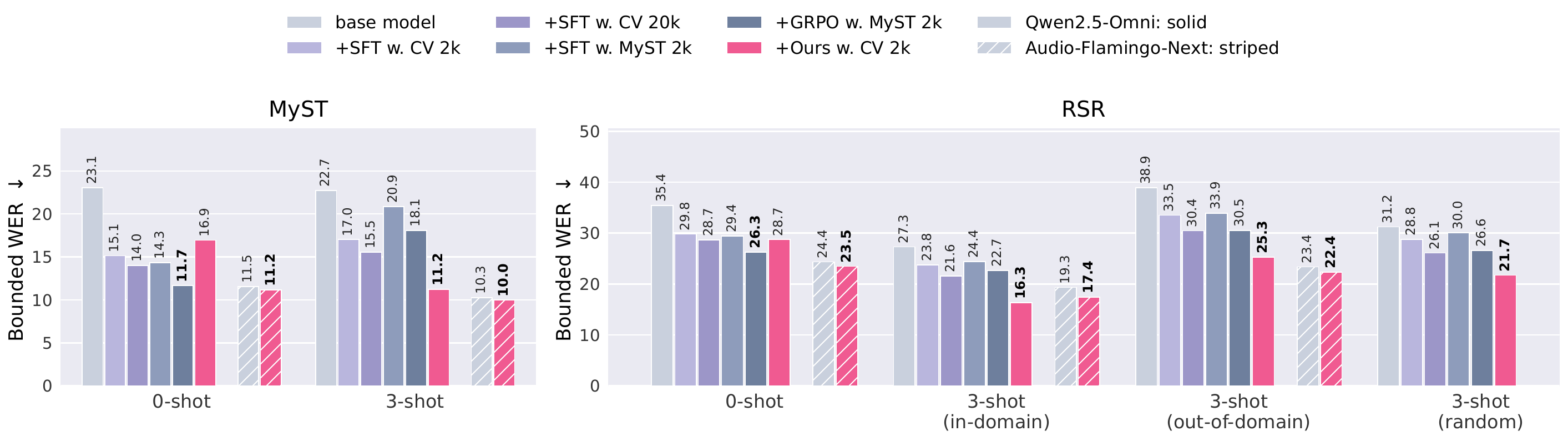}
\caption{
Child ASR: direct fine-tuning versus SICL-style GRPO, bounded WER (lower is better).
Solid bars represent $\textit{Qwen2.5-Omni}$, and striped bars represent \textit{Audio-Flamingo-Next}; gray bars indicate the untuned base model, and pink bars indicate FSA-GRPO.
Baselines differ only in training data and objective: SFT on 2k or 20k out-of-domain Common Voice, SFT on 2k related-domain MyST, or GRPO on that same MyST data; FSA-GRPO uses the 2k Common Voice budget with a demonstration-aware objective. All direct-tuning baselines are trained in zero-shot format; only FSA-GRPO is trained in a demonstration-conditioned format.
RSR demonstrations are drawn from the RSR train split (\textit{in-domain}) or MyST (\textit{out-of-domain}); the \textit{random} group studies sensitivity to demonstration quality by sampling uniformly from the RSR train split. Bold marks the best score within each group.
}
\label{fig:direct_ft_child_asr}
\end{figure*}

\paragraph{Performance Analysis.}

As shown in Figure~\ref{fig:direct_ft_child_asr}, the direct-tuning baselines reveal some useful patterns. First, under the same limited budget, related-domain adaptation is more effective than generic out-of-domain ASR tuning. Second, direct GRPO performs better than direct SFT in our setting, suggesting that RL-based post-training can extract stronger signal from limited adaptation data. Third, increasing the amount of out-of-domain Common Voice data partly compensates for domain mismatch, showing that larger generic ASR supervision can still improve transcription ability.

FSA-GRPO achieves the best few-shot performance whether the in-context examples are drawn from the target domain or from an out-of-domain dataset. This matches our design goal: FSA-GRPO teaches the model to better use demonstrations for low-resource adaptation. Notably, it substantially reduces RSR WER (by 53.9\% relative) without any in-domain training. We also observe that direct tuning can improve few-shot adaptation to some extent, but its gains remain limited compared with FSA-GRPO.

\paragraph{Sensitivity to Demonstration Quality.}

Both few-shot settings above use demonstrations retrieved by TICL~\cite{zheng2025ticltextembeddingknnspeech}, so neither isolates how much of the gain depends on retrieval. We add a third condition that replaces retrieval with uniform sampling from the RSR training pool, holding the pool, demonstration count, and prompt fixed and evaluating every model using the same pre-built demonstration set.

FSA-GRPO is the only model never hurt by its demonstrations. Every direct-tuning baseline performs worse than its own zero-shot result in at least one mismatched-demonstration condition, whereas FSA-GRPO improves in all three conditions and gains the most when the demonstrations are informative. The model thus learns to exploit demonstrations that carry signal and to fall back on its own transcription otherwise. Its spread across the three conditions is comparable to the stronger baselines; what differs is that the spread lies entirely below its zero-shot score.

\subsection{Transferability}

To examine whether the proposed recipe is specific to $\textit{Qwen2.5-Omni}$ or can transfer to other auditory LLMs, we further apply the same FSA-GRPO training recipe to \textit{Audio-Flamingo-Next}. This experiment keeps the same motivation as the Qwen experiments: the model is trained with the proposed few-shot ASR objective and evaluated using few-shot inference, without changing the downstream evaluation protocol.


As shown in Figure~\ref{fig:direct_ft_child_asr}, these results suggest that the proposed training recipe is not tied to a single model family. However, the gains on \textit{Audio-Flamingo-Next} are relatively modest compared with those on \textit{Qwen2.5-Omni}. One possible reason is that the GRPO configuration and reward hyperparameters were selected based on $\textit{Qwen2.5-Omni}$, while \textit{Audio-Flamingo-Next} may have a different optimal reward-weight sweet spot. Therefore, these results should be interpreted as preliminary evidence of cross-model transferability rather than a fully optimized result for \textit{Audio-Flamingo-Next}. Even under this non-tuned setting, the consistent improvements across MyST, RSR, and out-of-domain in-context evaluation indicate that FSA-GRPO can strengthen few-shot adaptation behavior across different auditory LLM backbones.

\begin{table}[t]
\centering
\small
\renewcommand{\arraystretch}{1.08}
\caption{Fixed-budget data-selection ablation for the RSR follow-up with the semantic-alignment reward recipe. Each selected subset contains 1k training samples. Here, \textit{ctx} denotes the quality of retrieved in-context examples, measured by similarity between the gold transcript and retrieved labels; \textit{perf} denotes the base model's current few-shot performance, measured by WER. Results are reported as bounded WER, where lower is better.}
\label{tab:ice_reward_rsr_followup}

\begin{tabularx}{\columnwidth}{@{}>{\raggedright\arraybackslash}X>{\centering\arraybackslash}X@{}}
\toprule
Training subset (1k) & Bounded WER \\
\midrule

\rowcolor{white}
$\textit{Qwen2.5-Omni}$
& 27.83 \\

\rowcolor{gray!15}
\textit{+ Random}
& 22.14 \\

\rowcolor{white}
\textit{+ Good ctx, good perf}
& 23.59 \\

\rowcolor{gray!15}
\textit{+ Bad ctx, bad perf}
& 21.37 \\

\rowcolor{white}
\textit{+ Good ctx, bad perf}
& \textbf{20.56} \\

\bottomrule
\end{tabularx}
\end{table}

\subsection{Data Selection Strategy}
\label{sec:data-selection}

\paragraph{Motivation.}

Since GRPO training is expensive, we conduct a fixed-budget ablation to identify which types of data instances provide the most useful training signal. Intuitively, examples on which the base model already performs well under the current few-shot prompting setting are unlikely to provide strong learning signal\footnote{Their rewards tend to be relatively flat because the model has already largely solved them.}. Therefore, we focus primarily on failure cases, where the model's current output still leaves room for improvement. However, in the few-shot setting, high WER can arise from two different sources: the retrieved in-context examples may contain useful signal, but the model may still fail to effectively use that context, or the retrieved examples may be poorly aligned and therefore confuse the model. These two cases should not be treated as equivalent. Our hypothesis is that good-context/bad-performance examples are more valuable for training.

  We construct 1k-example ablation subsets from the same few-shot training-instance pool described in Section~\ref{sec:training-data}. For each instance, we compute two diagnostic scores: context quality and the base model's current few-shot performance. Context quality is measured by \texttt{gold\_to\_ice\_best}, the maximum similarity between the gold transcript and the retrieved in-context labels. We define good context as \texttt{gold\_to\_ice\_best} $\geq 0.6$ and bad context otherwise. Current model performance is measured by the raw WER of the base few-shot prediction; we define bad performance as \texttt{raw\_wer} $\geq 0.2$ and good performance otherwise. This yields three diagnostic groups: good-context/good-performance, good-context/bad-performance, and bad-context/bad-performance, along with a random baseline. All groups are trained with the same GRPO setup and data budget, so differences can be attributed to the data-selection rule rather than training scale.

\paragraph{Ablation Results \& Analysis.}

Table~\ref{tab:ice_reward_rsr_followup} supports our data-selection hypothesis. The good-context/good-performance subset improves over the base model but underperforms the other selected subsets. In contrast, both bad-performance subsets outperform the random baseline. Among these failure cases, context quality further matters: good-context/bad-performance examples perform best, showing that useful retrieved demonstrations are important even when the base model initially fails to use them. Based on this result, we prioritize good-context/bad-performance examples in our main training recipe.

\begin{figure}[t]
    \centering
    \includegraphics[width=1\linewidth]{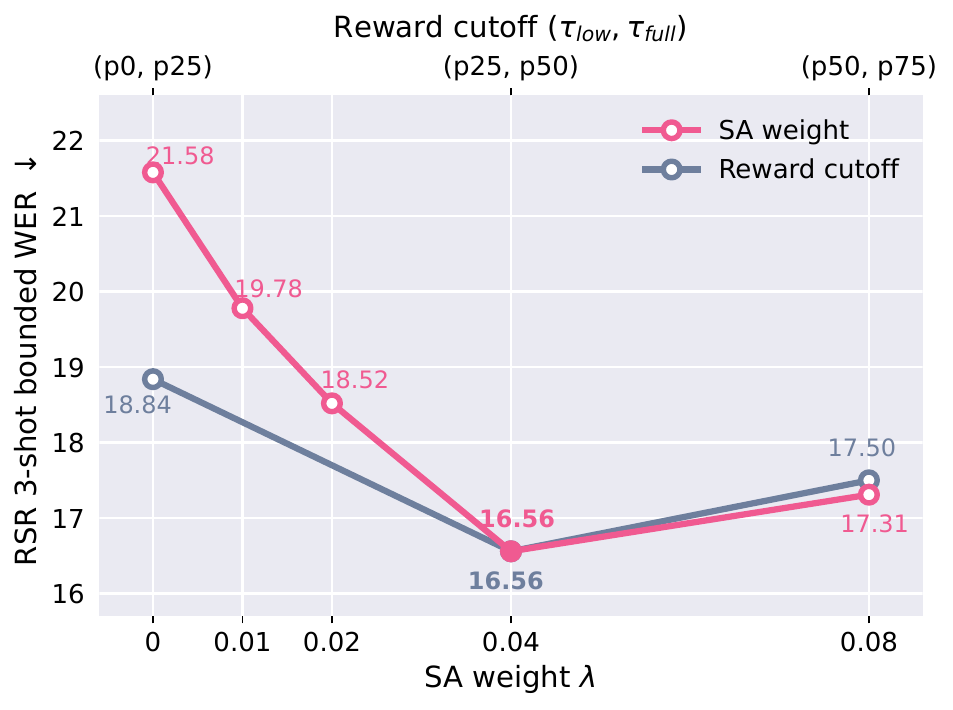}
\caption{
Ablation of the two semantic-alignment reward hyperparameters on RSR 3-shot evaluation with $\textit{Qwen2.5-Omni}$, reported as bounded WER (lower is better).
Each sweep varies one hyperparameter and holds the other at its chosen value: the SA weight sweep fixes the reward cutoff at $(\tau_{\mathrm{low}},\tau_{\mathrm{full}})=(p_{25},p_{50})$, and the cutoff sweep fixes $\lambda=0.04$; cutoffs are percentiles of the group-specific similarity distribution.
Training data, GRPO setup, and evaluation protocol are otherwise identical.
Filled markers denote the best setting of each sweep; both are the same run ($\lambda=0.04$ with the $(p25,p50)$ cutoff).
}
    \label{fig:sa_reward_weight_ablation}
    
\end{figure}

\subsection{Reward Ablation}
\label{sec:weight-ablation}

\paragraph{Reward Importance.}
After fixing the data-selection rule, we examine the importance of the auxiliary semantic-alignment reward. This experiment uses the same training data, model, GRPO configuration, reward preset, and RSR 3-shot evaluation protocol, and varies only the auxiliary reward weight $\lambda$. Intuitively, $\lambda$ controls how strongly the model is encouraged to align its predictions with the retrieved in-context examples. $\lambda=0$ corresponds to GRPO with only the WER reward, under which the model can improve transcription accuracy without necessarily learning to attend to the demonstrations. A small nonzero $\lambda$ provides a hint-like signal that encourages the model to leverage useful information from the in-context examples while still keeping ASR accuracy as the dominant objective. However, if $\lambda$ is too large, the model may overemphasize matching or copying the retrieved demonstrations, even when the query audio should remain the primary source of evidence.

\paragraph{Reward Cutoff.}
We then fix $\lambda=0.04$ and vary the reward cutoff $(\tau_{\mathrm{low}},\tau_{\mathrm{full}})$, the band over which the auxiliary reward ramps from zero to its full value. The upper cutoff is what keeps the term from rewarding verbatim copying: without it, reproducing a demonstration would be the easiest way to score. Cutoffs are percentiles of the group-specific similarity distribution, so the band adapts to how similar demonstrations typically are for a task. Both ends carry a risk: a band set too low is cleared by almost every generation and gives little signal, while a band set too high is reachable only by closely matching the demonstration labels, inviting the model to chase the reward by over-copying.

\paragraph{Ablation Results \& Analysis.}
Figure~\ref{fig:sa_reward_weight_ablation} reports both sweeps. The auxiliary reward consistently improves over the ASR-only control at every nonzero weight, with the best result at $\lambda=0.04$; performance degrades again at $\lambda=0.08$, indicating that the auxiliary reward is most effective as a moderate regularizing signal rather than a dominant optimization objective. The cutoff sweep has the same shape: the percentile-derived default performs best, and moving in either direction reduces performance. The two extremes degrade for different reasons. With the low band, the auxiliary reward saturates for nearly every generation and stops discriminating; with the high band, few generations clear it, so the term stays near zero and the run falls back to ASR-only behavior. Neither sweep is knife-edge sensitive, which supports setting both hyperparameters from the similarity distribution rather than tuning them per task.

\section{Conclusion}
This work introduces FSA-GRPO, an RL-based post-training recipe that teaches auditory LLMs to better use few-shot demonstrations during inference. To this end, we design a few-shot-aware reward that jointly encourages task correctness and effective use of the provided demonstrations. Notably, training with only English ASR data improves the model's general speech in-context learning ability across multiple held-out tasks, including child ASR, multilingual ASR, speech translation, and audio understanding. Through a controlled low-resource ASR evaluation, we further show that FSA-GRPO provides a more effective adaptation strategy than direct fine-tuning on related out-of-domain data. Overall, our results suggest that FSA-GRPO is a promising alternative for adapting auditory LLMs to low-resource tasks when in-domain training data are unavailable or limited.

\section*{Limitations}

This study has several limitations. First, although FSA-GRPO improves few-shot adaptation across multiple evaluation settings, most results are reported from a single fixed checkpoint with fixed decoding, retrieval, and demonstration settings. We do not report mean and standard deviation over multiple random seeds, so small differences between systems should be interpreted cautiously. Second, our main experiments are conducted on \textit{Qwen2.5-Omni}, and the transfer experiment on \textit{Audio-Flamingo-Next} uses the same recipe without model-specific hyperparameter tuning. Therefore, the Audio-Flamingo-Next results should be viewed as preliminary evidence of transferability rather than a fully optimized result for that model family.

Third, our training recipe is evaluated under a limited set of few-shot configurations. We mainly study a fixed demonstration budget and retrieval-based in-context examples, so the optimal behavior may change with different numbers of demonstrations, retrieval methods, prompt formats, or context lengths. Similarly, our data-selection rule, auxiliary reward weight, and similarity cutoffs are selected based on bounded-budget ablations; while they provide practical guidance for this setting, they may not be optimal for other auditory LLMs, languages, domains, or training budgets.

Fourth, our evaluation relies on benchmark datasets and automatic metrics, including WER, bounded WER, BLEU, and accuracy. These metrics are useful for controlled comparison, but they do not fully capture all types of speech understanding errors, user experience, or downstream consequences. In addition, we apply text normalization before computing WER and BLEU, which reduces surface-form variation but may also hide some punctuation, casing, or formatting differences. Finally, although child ASR is used as a motivating low-resource setting, our results do not establish readiness for clinical, educational placement, legal, or other high-stakes deployment. Real-world use would require additional evaluation across speaker groups, recording conditions, and application-specific safety requirements.

\section*{Data Use and Privacy}

All experiments use existing datasets under their original licenses and access conditions. Some contain sensitive speech, including recordings of children and multilingual or accented speakers. We collect no new human-subject data and rely on providers' consent, access-control, and de-identification procedures. We do not attempt speaker identification, attribute inference, or identity linkage, and report only aggregate metrics. Any released code or derived artifacts will exclude materials that could re-identify participants unless already public under the original terms.

\section*{Potential Risks and Mitigations}

Improved speech recognition can support accessibility, education, and low-resource language technologies, but may also enable surveillance, profiling, or unsupported high-stakes use. These risks are acute because ASR errors often disproportionately affect children, accented speakers, and low-resource languages. We therefore report results across diverse settings, do not treat benchmark gains as evidence of readiness for rights-impacting decisions, and recommend informed consent, data-security controls, human oversight, and monitoring of differential error rates in any deployment.

\section*{Misuse Considerations}

FSA-GRPO lowers the barrier to adapting general speech models with limited data, which benefits low-resource research but may also enable misuse in new domains. We restrict our claims to research evaluation and recommend documenting training data, intended use, retrieval settings, reward design, and known failure modes. The resulting models do not provide clinical, diagnostic, or professional judgments and should not replace qualified experts.

\section*{Environmental Impact}

GRPO is compute-intensive because it samples multiple outputs during training. We limit its footprint through parameter-efficient adaptation, bounded training budgets, fixed evaluation settings, and ablations that improve sample efficiency. We report key training configurations to support reproducibility and future compute estimates.

\section*{AI Usage Statement}

Generative AI tools assisted only with grammar, clarity, organization, and \LaTeX{} formatting. The authors produced and verified all research ideas, experimental design, implementation, data processing, results, analyses, and claims, and take full responsibility for the manuscript. No private, restricted, or unpublished dataset content was provided beyond manuscript text, and the tools did not generate or modify evaluation data, labels, model outputs, or reported metrics.

\section*{Software and Packages}

Experiments use open-source toolchains including PyTorch, Hugging Face Transformers and Datasets, LoRA libraries, and standard evaluation packages. We use official benchmark protocols or public evaluation scripts when available.

\section*{Descriptive Statistics}

We report corpus-level WER, bounded WER, BLEU, and accuracy on the full evaluation sets. Unless noted, each result comes from one run of a fixed checkpoint with fixed decoding, retrieval, and demonstration settings. We do not report variation across random seeds; small differences should therefore be interpreted cautiously, with emphasis on larger trends consistent across tasks, shot settings, and model backbones.

\section*{Acknowledgments}
This material is based upon work supported under the AI Research Institutes program by National Science Foundation and the Institute of Education Sciences, U.S. Department of Education, through Award \#2229873 - National AI Institute for Exceptional Education. Any opinions, findings and conclusions or recommendations expressed in this material are those of the author(s) and do not necessarily reflect the views of the National Science Foundation, the Institute of Education Sciences, or the U.S. Department of Education.
This work used the Delta system at the National Center for Supercomputing Applications through allocation beiq-delta-gpu from the Advanced Cyberinfrastructure Coordination Ecosystem: Services \& Support (ACCESS) program, which is supported by National Science Foundation grants \#2138259, \#2138286, \#2138307, \#2137603, and \#2138296.


\bibliography{custom}

@article{wang2026towards,
  title={Towards general auditory intelligence for machine listening and speaking},
  author={Wang, Siyin and Jin, Zengrui and Tang, Changli and Li, Qiujia and Li, Bo and Chen, Chen and Hu, Yuchen and others},
  journal={Nature Machine Intelligence},
  pages={1--20},
  year={2026},
  publisher={Nature Publishing Group UK London}
}

@inproceedings{min2022metaicl,
  title={Metaicl: Learning to learn in context},
  author={Min, Sewon and Lewis, Mike and Zettlemoyer, Luke and Hajishirzi, Hannaneh},
  booktitle={Proceedings of the 2022 conference of the North American chapter of the Association for Computational Linguistics: Human Language Technologies},
  pages={2791--2809},
  year={2022}
}

@inproceedings{zheng2025ticltextembeddingknnspeech,
  title={{TICL: Text-Embedding KNN for Speech In-Context Learning Unlocks Speech Recognition Abilities of Large Multimodal Models}},
  author={Zheng, Haolong and Yegorova, Yekaterina and Hasegawa-Johnson, Mark},
  booktitle={ICASSP 2026 -- 2026 IEEE International Conference on Acoustics, Speech and Signal Processing (ICASSP)},
  pages={17912--17916},
  year={2026},
  doi={10.1109/ICASSP55912.2026.11461434}
}

@inproceedings{zhou2025m2r,
  title={{M2R-Whisper: Multi-stage and Multi-scale Retrieval Augmentation for Enhancing Whisper}},
  author={Zhou, Jiaming and Zhao, Shiwan and He, Jiabei and Wang, Hui and Zeng, Wenjia and Chen, Yong and Sun, Haoqin and Kong, Aobo and Qin, Yong},
  booktitle={ICASSP},
  pages={1--5},
  year={2025},
  organization={IEEE}
}

@inproceedings{wang2024can,
  title={{Can Whisper Perform Speech-Based In-Context Learning?}},
  author={Wang, Siyin and Yang, Chao-Han and Wu, Ji and Zhang, Chao},
  booktitle={ICASSP},
  pages={13421--13425},
  year={2024},
  organization={IEEE}
}

@misc{zheng2025ticlcasestudyspeech,
      title={{TICL+: A Case Study On Speech In-Context Learning for Children's Speech Recognition}}, 
      author={Haolong Zheng and Yekaterina Yegorova and Mark Hasegawa-Johnson},
      year={2025},
      eprint={2512.18263},
      archivePrefix={arXiv},
      primaryClass={eess.AS},
      url={https://arxiv.org/abs/2512.18263}, 
}

@misc{coreteam2025mimoaudio,
      title={{MiMo-Audio: Audio Language Models are Few-Shot Learners}}, 
      author={LLM-Core-Team Xiaomi},
      year={2025},
      url={https://github.com/XiaomiMiMo/MiMo-Audio}, 
}

@inproceedings{brown2020language,
 author = {Brown, Tom and Mann, Benjamin and Ryder, Nick and Subbiah, Melanie and Kaplan, Jared D and Dhariwal, Prafulla and Neelakantan, Arvind and Shyam, Pranav and Sastry, Girish and Askell, Amanda and others},
 booktitle = {Advances in Neural Information Processing Systems},
 pages = {1877--1901},
 title = {{Language Models are Few-Shot Learners}},
 volume = {33},
 year = {2020}
}

@inproceedings{huang2023language,
  title={{Language Is Not All You Need: Aligning Perception with Language Models}},
  author={Huang, Shaohan and Dong, Li and Wang, Wenhui and Hao, Yaru and Singhal, Saksham and Ma, Shuming and Lv, Tengchao and Cui, Lei and Mohammed, Owais Khan and Patra, Barun and others},
  booktitle={Advances in Neural Information Processing Systems},
  volume={36},
  pages={72096--72109},
  year={2023}
}

@article{omnilingual2025omnilingual,
  title={{Omnilingual ASR: Open-Source Multilingual Speech Recognition for 1600+ Languages}},
  author={Keren, Gil and Kozhevnikov, Artyom and Meng, Yen and Ropers, Christophe and Setzler, Matthew and Wang, Skyler and Adebara, Ife and Auli, Michael and Balioglu, Can and others},
  journal={arXiv preprint arXiv:2511.09690},
  year={2025}
}

@inproceedings{chen2024salm,
  title={{SALM: Speech-augmented Language Model with In-context Learning for Speech Recognition and Translation}},
  author={Chen, Zhehuai and Huang, He and Andrusenko, Andrei and Hrinchuk, Oleksii and Puvvada, Krishna C and Li, Jason and Ghosh, Subhankar and Balam, Jagadeesh and Ginsburg, Boris},
  booktitle={ICASSP},
  pages={13521--13525},
  year={2024},
  organization={IEEE}
}

@inproceedings{kong2024audio,
  title={{Audio Flamingo: A Novel Audio Language Model with Few-Shot Learning and Dialogue Abilities}},
  author={Kong, Zhifeng and Goel, Arushi and Badlani, Rohan and Ping, Wei and Valle, Rafael and Catanzaro, Bryan},
  booktitle={Proceedings of the 41st International Conference on Machine Learning},
  pages={25125--25148},
  year={2024}
}

@inproceedings{ihori2025few,
  author={Ihori, Mana and Yamane, Taiga and Kawata, Naotaka and Makishima, Naoki and Tanaka, Tomohiro and Suzuki, Satoshi and Orihashi, Shota and Masumura, Ryo},
  booktitle={ASRU}, 
  title={{Few-shot Personalization via In-Context Learning for Speech Emotion Recognition based on Speech-Language Model}}, 
  year={2025},
  volume={},
  number={},
  pages={1-6},
}

@inproceedings{hsu2024smile,
  title={{SMILE}: Speech Meta In-Context Learning for Low-Resource Language Automatic Speech Recognition},
  author={Hsu, Ming-Hao and Lee, Hung-yi},
  booktitle={2025 IEEE Automatic Speech Recognition and Understanding Workshop (ASRU)},
  pages={1--7},
  year={2025},
  doi={10.1109/ASRU65441.2025.11434760}
}

@inproceedings{wang2020covost,
  title     = {CoVoST 2 and Massively Multilingual Speech Translation},
  author    = {Changhan Wang and Anne Wu and Jiatao Gu and Juan Pino},
  year      = {2021},
  booktitle = {INTERSPEECH},
  pages     = {2247--2251},
  doi       = {10.21437/Interspeech.2021-2027},
  issn      = {2958-1796},
}

@inproceedings{commonvoice:2020,
  author = {Ardila, R. and Branson, M. and Davis, K. and Henretty, M. and Kohler, M. and Meyer, J. and Morais, R. and Saunders, L. and Tyers, F. M. and Weber, G.},
  title = {{Common Voice: A Massively-Multilingual Speech Corpus}},
  booktitle = {Proceedings of the 12th Conference on Language Resources and Evaluation},
  pages = {4218--4222},
  year = {2020}
}

@article{Qwen2.5-Omni,
  title={{Qwen2.5-Omni Technical Report}},
  author={Jin, Xu and Zhifang, Guo and Jinzheng, He and Hangrui, Hu and Ting, He and Shuai, Bai and Keqin, Chen and Jialin, Wang and Yang, Fan and Kai, Dang and Bin, Zhang and Xiong, Wang and Yunfei, Chu and Junyang, Lin},
  journal={arXiv preprint arXiv:2503.20215},
  year={2025}
}

@inproceedings{zheng2025interspeech,
  title     = {{The Interspeech 2025 Speech Accessibility Project Challenge}},
  author    = {Xiuwen Zheng and Bornali Phukon and Jonghwan Na and Ed Cutrell and Kyu J. Han and Mark Hasegawa-Johnson and Pan-Pan Jiang and Aadhrik Kuila and Colin Lea and Bob MacDonald and Gautam Mantena and Venkatesh Ravichandran and Leda Sari and Katrin Tomanek and Chang D. Yoo and Chris Zwilling},
  year      = {2025},
  booktitle = {{INTERSPEECH}},
  pages     = {3269--3273},
  doi       = {10.21437/Interspeech.2025-566},
  issn      = {2958-1796},
}

@inproceedings{pradhan2024my,
  title={{My Science Tutor (MyST) -- A Large Corpus of Children's Conversational Speech}},
  author={Pradhan, Sameer and Cole, Ronald and Ward, Wayne},
  booktitle={Proceedings of the 2024 Joint International Conference on Computational Linguistics, Language Resources and Evaluation},
  pages={12040--12045},
  year={2024}
}

@article{ai4exceptionaled_rsr_hf,
  title={{Diagnostic Accuracy of Sentence Recall and Past Tense Measures for Identifying Children's Language Impairments}},
  author={Redmond, Sean M and Ash, Andrea C and Christopulos, Tyler T and Pfaff, Theresa},
  journal={Journal of Speech, Language, and Hearing Research},
  volume={62},
  number={7},
  pages={2438--2454},
  year={2019},
  publisher={American Speech-Language-Hearing Association}
}

@article{yang2024uniaudio,
  title={Uniaudio 1.5: Large language model-driven audio codec is a few-shot audio task learner},
  author={Yang, Dongchao and Guo, Haohan and Wang, Yuanyuan and Huang, Rongjie and Li, Xiang and Tan, Xu and Wu, Xixin and Meng, Helen},
  journal={NeurIPS},
  pages={56802--56827},
  year={2024}
}

@inproceedings{pan2023cosmic,
  title={{COSMIC}: Data Efficient Instruction-tuning For Speech In-Context Learning},
  author={Pan, Jing and Wu, Jian and Gaur, Yashesh and Sivasankaran, Sunit and Chen, Zhuo and Liu, Shujie and Li, Jinyu},
  booktitle={Interspeech 2024},
  pages={4164--4168},
  year={2024},
  doi={10.21437/Interspeech.2024-1346}
}

@inproceedings{wang2024bayesian,
  title={Bayesian Example Selection Improves In-Context Learning for Speech, Text and Visual Modalities},
  author={Wang, Siyin and Yang, Chao-Han and Wu, Ji and Zhang, Chao},
  booktitle={EMNLP},
  pages={20812--20828},
  year={2024}
}

@misc{zheng2026metasiclglobalizingauditoryllms,
  title         = {{MetaSICL: Globalizing Auditory LLMs for Underserved Speakers and Languages via Meta Speech In-Context Learning}},
  author        = {Zheng, Haolong and Wang, Siyin and Jin, Zengrui and Hasegawa-Johnson, Mark},
  year          = {2026},
  eprint        = {2601.18904},
  archivePrefix = {arXiv},
  primaryClass  = {cs.SD},
  url           = {https://arxiv.org/abs/2601.18904}
}

@article{ghosh2026audio,
  title={Audio Flamingo Next: Next-Generation Open Audio-Language Models for Speech, Sound, and Music},
  author={Ghosh, Sreyan and Goel, Arushi and Jayakumar, Kaousheik and Koroshinadze, Lasha and Anand, Nishit and Kong, Zhifeng and Gururani, Siddharth and Lee, Sang-gil and Kim, Jaehyeon and Aljafari, Aya and others},
  journal={arXiv preprint arXiv:2604.10905},
  year={2026}
}

@inproceedings{sakshi2025mmau,
  title={Mmau: A massive multi-task audio understanding and reasoning benchmark},
  author={Sakshi, Sakshi and Tyagi, Utkarsh and Kumar, Sonal and Seth, Ashish and Selvakumar, Ramaneswaran and Nieto, Oriol and Duraiswami, Ramani and Ghosh, Sreyan and Manocha, Dinesh},
  booktitle={International Conference on Learning Representations},
  volume={2025},
  pages={84929--84964},
  year={2025}
}

@article{ma2026mmar,
  title={Mmar: A challenging benchmark for deep reasoning in speech, audio, music, and their mix},
  author={Ma, Ziyang and Ma, Yinghao and Zhu, Yanqiao and Yang, Chen and Chao, Yi-Wen and Xu, Ruiyang and Chen, Wenxi and Chen, Yuanzhe and Chen, Zhuo and Cong, Jian and others},
  journal={Advances in Neural Information Processing Systems},
  volume={38},
  year={2026}
}

@inproceedings{radford2023robust,
  title={Robust Speech Recognition via Large-Scale Weak Supervision},
  author={Radford, Alec and Kim, Jong Wook and Xu, Tao and Brockman, Greg and McLeavey, Christine and Sutskever, Ilya},
  booktitle={Proceedings of the 40th International Conference on Machine Learning},
  pages={28492--28518},
  year={2023},
  organization={PMLR}
}

@article{shao2024deepseekmath,
  title={Deepseekmath: Pushing the limits of mathematical reasoning in open language models},
  author={Shao, Zhihong and Wang, Peiyi and Zhu, Qihao and Xu, Runxin and Song, Junxiao and Bi, Xiao and Zhang, Haowei and Zhang, Mingchuan and Li, YK and Wu, Yang and others},
  journal={arXiv preprint arXiv:2402.03300},
  year={2024}
}

@article{schulman2017proximal,
  title={Proximal policy optimization algorithms},
  author={Schulman, John and Wolski, Filip and Dhariwal, Prafulla and Radford, Alec and Klimov, Oleg},
  journal={arXiv preprint arXiv:1707.06347},
  year={2017}
}

@article{guo2025deepseekr1,
  title={{DeepSeek-R1} incentivizes reasoning in {LLMs} through reinforcement learning},
  author={Guo, Daya and Yang, Dejian and Zhang, Haowei and Song, Junxiao and Wang, Peiyi and Zhu, Qihao and Xu, Runxin and Zhang, Ruoyu and Ma, Shirong and Bi, Xiao and others},
  journal={Nature},
  volume={645},
  number={8081},
  pages={633--638},
  year={2025},
  doi={10.1038/s41586-025-09422-z}
}

@inproceedings{hu2022lora,
  title={LoRA: Low-Rank Adaptation of Large Language Models},
  author={Hu, Edward J. and Shen, Yelong and Wallis, Phillip and Allen-Zhu, Zeyuan and Li, Yuanzhi and Wang, Shean and Wang, Lu and Chen, Weizhu},
  booktitle={International Conference on Learning Representations},
  year={2022}
}

@inproceedings{wang2026end,
  title={End-to-end Listen, Look, Speak and Act},
  author={Wang, Siyin and Yu, Wenyi and Chen, Xianzhao and Tian, Xiaohai and Zhang, Jun and Lu, Lu and Zhang, Chao},
  booktitle={Proc. ICLR},
  year={2026},
  address={Rio de Janeiro}
}

@inproceedings{fong25_interspeech,
  title     = {{Speech LLMs in Low-Resource Scenarios: Data Volume Requirements and the Impact of Pretraining on High-Resource Languages}},
  author    = {Seraphina Fong and Marco Matassoni and Alessio Brutti},
  year      = {2025},
  booktitle = {{Interspeech 2025}},
  pages     = {2003--2007},
  doi       = {10.21437/Interspeech.2025-764},
  issn      = {2958-1796},
}

@article{ghai2010exploring,
  title={Exploring the effect of differences in the acoustic correlates of adults' and children's speech in the context of automatic speech recognition},
  author={Ghai, Shweta and Sinha, Rohit},
  journal={EURASIP Journal on Audio, Speech, and Music Processing},
  volume={2010},
  number={1},
  pages={318785},
  year={2010},
  publisher={Springer}
}

@inproceedings{gerosa2009review,
  title={A review of ASR technologies for children's speech},
  author={Gerosa, Matteo and Giuliani, Diego and Narayanan, Shrikanth and Potamianos, Alexandros},
  booktitle={Proceedings of the 2nd Workshop on Child, Computer and Interaction},
  pages={1--8},
  year={2009}
}

@article{shivakumar2020transfer,
  title={Transfer learning from adult to children for speech recognition: Evaluation, analysis and recommendations},
  author={Shivakumar, Prashanth Gurunath and Georgiou, Panayiotis},
  journal={Computer speech \& language},
  volume={63},
  pages={101077},
  year={2020},
  publisher={Elsevier}
}

@inproceedings{fan2024benchmarking,
  title={Benchmarking Children's {ASR} with Supervised and Self-supervised Speech Foundation Models},
  author={Fan, Ruchao and Shankar, Natarajan Balaji and Alwan, Abeer},
  booktitle={Interspeech 2024},
  pages={5173--5177},
  year={2024},
  doi={10.21437/Interspeech.2024-1353}
}

@article{li2025reinforcement,
  title={Reinforcement learning outperforms supervised fine-tuning: A case study on audio question answering},
  author={Li, Gang and Liu, Jizhong and Dinkel, Heinrich and Niu, Yadong and Zhang, Junbo and Luan, Jian},
  journal={arXiv preprint arXiv:2503.11197},
  year={2025}
}

@inproceedings{elmakies2025advancing,
  title={Advancing Speech Understanding in Speech-Aware Language Models with {GRPO}},
  author={Elmakies, Avishai and Aronowitz, Hagai and Shabtay, Nimrod and Schwartz, Eli and Hoory, Ron and Dekel, Avihu},
  booktitle={ICASSP 2026 -- 2026 IEEE International Conference on Acoustics, Speech and Signal Processing (ICASSP)},
  pages={17797--17801},
  year={2026},
  doi={10.1109/ICASSP55912.2026.11463105}
}

@inproceedings{shivakumar2025grpospeech,
  title={Group Relative Policy Optimization for Speech Recognition},
  author={Shivakumar, Prashanth Gurunath and Gu, Yile and Gandhe, Ankur and Bulyko, Ivan},
  booktitle={2025 IEEE Automatic Speech Recognition and Understanding Workshop (ASRU)},
  pages={1--7},
  year={2025},
  doi={10.1109/ASRU65441.2025.11434657}
}

@inproceedings{reimers-2019-sentence-bert,
    title = "Sentence-{BERT}: Sentence Embeddings using Siamese {BERT}-Networks",
    author = "Reimers, Nils and Gurevych, Iryna",
    booktitle = "Proceedings of the 2019 Conference on Empirical Methods in Natural Language Processing and the 9th International Joint Conference on Natural Language Processing",
    year = "2019",
    pages = "3982--3992",
    doi = "10.18653/v1/D19-1410"
}

@inproceedings{chang2024exploring,
  title     = {{Exploring In-Context Learning of Textless Speech Language Model for Speech Classification Tasks}},
  author    = {Chang, Kai-Wei and Hsu, Ming-Hao and Li, Shan-Wen and Lee, Hung-yi},
  year      = {2024},
  booktitle = {{Interspeech 2024}},
  pages     = {4139--4143},
  doi       = {10.21437/Interspeech.2024-1932},
  issn      = {2958-1796}
}

@inproceedings{agrawal2025spoken,
  title     = {{Spoken Language Understanding on Unseen Tasks With In-Context Learning}},
  author    = {Agrawal, Neeraj and Ganapathy, Sriram},
  year      = {2025},
  booktitle = {{Interspeech 2025}},
  pages     = {4103--4107},
  doi       = {10.21437/Interspeech.2025-1467},
  issn      = {2958-1796}
}

@inproceedings{li2026multimodal,
  title={Multimodal In-context Learning for {ASR} of Low-resource Languages},
  author={Li, Zhaolin and Niehues, Jan},
  booktitle={Findings of the Association for Computational Linguistics: ACL 2026},
  pages={24745--24760},
  year={2026},
  doi={10.18653/v1/2026.findings-acl.1239}
}

@inproceedings{joshi-etal-2020-state,
  title     = {The State and Fate of Linguistic Diversity and Inclusion in the {NLP} World},
  author    = {Joshi, Pratik and Santy, Sebastin and Budhiraja, Amar and Bali, Kalika and Choudhury, Monojit},
  booktitle = {Proceedings of the 58th Annual Meeting of the Association for Computational Linguistics},
  year      = {2020},
  pages     = {6282--6293},
  publisher = {Association for Computational Linguistics},
  doi       = {10.18653/v1/2020.acl-main.560},
  url       = {https://aclanthology.org/2020.acl-main.560}
}

@inproceedings{rei-etal-2020-comet,
  title     = {{COMET}: A Neural Framework for {MT} Evaluation},
  author    = {Rei, Ricardo and Stewart, Craig and Farinha, Ana C and Lavie, Alon},
  booktitle = {Proceedings of the 2020 Conference on Empirical Methods in Natural Language Processing (EMNLP)},
  year      = {2020},
  pages     = {2685--2702},
  publisher = {Association for Computational Linguistics},
  doi       = {10.18653/v1/2020.emnlp-main.213},
  url       = {https://aclanthology.org/2020.emnlp-main.213}
}

@inproceedings{rei-etal-2022-comet,
  title     = {{COMET}-22: Unbabel-{IST} 2022 Submission for the Metrics Shared Task},
  author    = {Rei, Ricardo and C. de Souza, Jos{\'e} G. and Alves, Duarte and Zerva, Chrysoula and Farinha, Ana C and Glushkova, Taisiya and Lavie, Alon and Coheur, Luisa and Martins, Andr{\'e} F. T.},
  booktitle = {Proceedings of the Seventh Conference on Machine Translation (WMT)},
  year      = {2022},
  pages     = {578--585},
  publisher = {Association for Computational Linguistics},
  doi       = {10.18653/v1/2022.wmt-1.52},
  url       = {https://aclanthology.org/2022.wmt-1.52}
}

@inproceedings{conneau2023fleurs,
  title     = {{FLEURS}: Few-Shot Learning Evaluation of Universal Representations of Speech},
  author    = {Conneau, Alexis and Ma, Min and Khanuja, Simran and Zhang, Yu and Axelrod, Vera and Dalmia, Siddharth and Riesa, Jason and Rivera, Clara and Bapna, Ankur},
  booktitle = {2022 IEEE Spoken Language Technology Workshop (SLT)},
  year      = {2023},
  pages     = {798--805},
  doi       = {10.1109/SLT54892.2023.10023141}
}

@inproceedings{pratap2020mls,
  title     = {{MLS}: A Large-Scale Multilingual Dataset for Speech Research},
  author    = {Pratap, Vineel and Xu, Qiantong and Sriram, Anuroop and Synnaeve, Gabriel and Collobert, Ronan},
  booktitle = {Interspeech 2020},
  year      = {2020},
  pages     = {2757--2761},
  doi       = {10.21437/Interspeech.2020-2826},
  issn      = {2958-1796}
}

@misc{srivastav2025openasr,
  title         = {{Open ASR Leaderboard: Towards Reproducible and Transparent Multilingual and Long-Form Speech Recognition Evaluation}},
  author        = {Srivastav, Vaibhav and Zheng, Steven and Bezzam, Eric and Le Bihan, Eustache and Koluguri, Nithin Rao and {\.Z}elasko, Piotr and Majumdar, Somshubra and Moumen, Adel and Gandhi, Sanchit},
  year          = {2025},
  eprint        = {2510.06961},
  archivePrefix = {arXiv},
  primaryClass  = {eess.AS},
  url           = {https://arxiv.org/abs/2510.06961}
}

@inproceedings{papineni-etal-2002-bleu,
  title     = {{BLEU}: a Method for Automatic Evaluation of Machine Translation},
  author    = {Papineni, Kishore and Roukos, Salim and Ward, Todd and Zhu, Wei-Jing},
  booktitle = {Proceedings of the 40th Annual Meeting of the Association for Computational Linguistics},
  year      = {2002},
  pages     = {311--318},
  publisher = {Association for Computational Linguistics},
  doi       = {10.3115/1073083.1073135},
  url       = {https://aclanthology.org/P02-1040}
}

@inproceedings{post-2018-call,
  title     = {A Call for Clarity in Reporting {BLEU} Scores},
  author    = {Post, Matt},
  booktitle = {Proceedings of the Third Conference on Machine Translation: Research Papers},
  year      = {2018},
  pages     = {186--191},
  publisher = {Association for Computational Linguistics},
  doi       = {10.18653/v1/W18-6319},
  url       = {https://aclanthology.org/W18-6319}
}

\appendix

\section{FSA-GRPO Algorithm}
\begin{algorithm}[h]
\small
\caption{FSA-GRPO}
\label{alg:training-overview}
\DontPrintSemicolon

\KwIn{
Query set $\mathcal{D}_{\mathrm{query}}$,
demonstration pool $\mathcal{D}_{\mathrm{pool}}$,
retriever $\mathcal{R}$,
policy $\pi_{\theta}$,
reference model $\pi_{\mathrm{ref}}$,
number of demonstrations $k$,
group size $G$,
SA reward weight $\lambda$
}

\For{$step \gets 1$ \KwTo $\textit{Total\_Step}$}{
Sample $(x_q,y_q^\star)\sim \mathcal{D}_{\mathrm{query}}$\;

Retrieve demonstrations
$\mathcal{I}_q=\{(x_j,y_j)\}_{j=1}^{k}
\leftarrow \mathcal{R}(x_q,\mathcal{D}_{\mathrm{pool}})$\;

Construct few-shot prompt
$p_q=[x_1,y_1,\ldots,x_k,y_k,x_q]$\;

Sample candidate responses
$\{y_i\}_{i=1}^{G}\sim \pi_{\theta_{\mathrm{old}}}(\cdot \mid p_q)$\;

Compute rewards
$r_{\mathrm{ASR}}(x_q,y_i)$ and $r_{\mathrm{SA}}(x_q,y_i)$
using Eq.~\ref{eq:asr_reward} and Eq.~\ref{eq:sa_reward}\;

Combine rewards into
$\mathbb{R}_i=r_{\mathrm{ASR}}(x_q,y_i)+\lambda r_{\mathrm{SA}}(x_q,y_i)$
following Eq.~\ref{eq:final_reward}\;

Normalize rewards within the group to obtain
$\{\hat{A}_i\}_{i=1}^{G}$ using Eq.~\ref{eq:grpo_advantage}\;

Update $\pi_{\theta}$ with the GRPO objective in Eq.~\ref{eq:grpo_objective},
using reference model $\pi_{\mathrm{ref}}$\;
}
\end{algorithm}

\section{GRPO Objective}
\label{app:grpo}

For each few-shot prompt $p_q$, the old policy samples a group of $G$ responses
$\{y_i\}_{i=1}^{G}\sim \pi_{\theta_{\mathrm{old}}}(\cdot \mid p_q)$.
Each response receives the FSA-GRPO reward defined in Eq.~\ref{eq:final_reward},
and the group-normalized advantage is computed following Eq.~\ref{eq:grpo_advantage}.

We optimize the policy with the clipped GRPO objective in Eq.~\ref{eq:grpo_objective}.
\begin{figure*}[t]
\begin{equation}
\mathcal{L}_{\mathrm{GRPO}}(\theta)
= -\frac{1}{G}\sum_{i=1}^{G}\frac{1}{T_i}\sum_{t=1}^{T_i}
\Big[
\min \big( \rho_{i,t}(\theta)\hat{A}_i,\;
\operatorname{clip}(\rho_{i,t}(\theta),1-\epsilon,1+\epsilon)\hat{A}_i \big)
- \beta D_{\mathrm{KL}}^{i,t}
\Big],
\label{eq:grpo_objective}
\end{equation}
\end{figure*}
The importance ratio is
\begin{equation}
\rho_{i,t}(\theta)
=
\frac{
\pi_{\theta}(y_{i,t}\mid p_q,y_{i,<t})
}{
\pi_{\theta_{\mathrm{old}}}(y_{i,t}\mid p_q,y_{i,<t})
}.
\label{eq:grpo_ratio}
\end{equation}

The KL penalty is computed against the reference model:
\begin{equation}
\begin{aligned}
D_{\mathrm{KL}}^{i,t}
=
&\frac{
\pi_{\mathrm{ref}}(y_{i,t}\mid p_q,y_{i,<t})
}{
\pi_{\theta}(y_{i,t}\mid p_q,y_{i,<t})
}
\\
&-
\log
\frac{
\pi_{\mathrm{ref}}(y_{i,t}\mid p_q,y_{i,<t})
}{
\pi_{\theta}(y_{i,t}\mid p_q,y_{i,<t})
}
-1.
\end{aligned}
\label{eq:grpo_kl}
\end{equation}

\section{Evaluation Data}
\label{appendix:eval-data}

\paragraph{Child ASR.}
MyST~\cite{pradhan2024my} consists of spontaneous speech from students in Grades 3 to 5, while RSR~\cite{ai4exceptionaled_rsr_hf} contains scripted sentence-recall recordings from children aged 5 to 9.

\paragraph{Audio Understanding and Reasoning.}
MMAU~\cite{sakshi2025mmau} evaluates a wide range of audio skills, emphasizing perception and domain-specific reasoning. MMAR~\cite{ma2026mmar} further covers speaker, environment, and content reasoning, as well as audio quality comparison, music understanding, anomaly detection, spatial and temporal reasoning, and general reasoning. For both we report accuracy on the public test split using the official evaluation scripts.

\paragraph{Multilingual ASR and Speech Translation.}
From Common Voice~\cite{commonvoice:2020} we use the German, French, and Chinese subsets, and from CoVoST2~\cite{wang2020covost} the English--Japanese pair in both directions; the five additional directions are English to German and Chinese, and German, Chinese, and Portuguese to English. The Open ASR leaderboard~\cite{srivastav2025openasr} multilingual sets cover Common Voice in six languages and MLS~\cite{pratap2020mls} in six. For FLEURS~\cite{conneau2023fleurs} we evaluate every language for which a same-language training split is available as a retrieval pool, and group languages by the resource classes of \citet{joshi-etal-2020-state}.

\section{Full Result Tables}
\label{appendix:full-results}

\begin{table*}[t]
\centering
\small
\setlength{\tabcolsep}{4pt}
\caption{Full OpenASR-multilingual results (bounded WER, lower is better). \texttt{mcv} = Common Voice, \texttt{mls} = Multilingual LibriSpeech. Bold marks the best score per shot setting.}
\label{tab:openasr_full}
\begin{tabular}{l*{8}{c}}
\toprule
 & \multicolumn{2}{c}{Qwen2.5-Omni} & \multicolumn{2}{c}{+MetaSICL} & \multicolumn{2}{c}{+SFT w. same data} & \multicolumn{2}{c}{+FSA-GRPO (Ours)} \\
\cmidrule(lr){2-3}\cmidrule(lr){4-5}\cmidrule(lr){6-7}\cmidrule(lr){8-9}
Config & 0-shot & 3-shot & 0-shot & 3-shot & 0-shot & 3-shot & 0-shot & 3-shot \\
\midrule
mcv\_de & 8.10 & 8.68 & 8.31 & 8.12 & 9.03 & 9.03 & \textbf{6.49} & \textbf{6.10} \\
mcv\_en & 8.56 & 9.36 & 8.16 & 8.13 & 9.84 & 9.45 & \textbf{6.10} & \textbf{5.84} \\
mcv\_es & 5.83 & 6.23 & 5.99 & 5.79 & 6.77 & 7.07 & \textbf{4.54} & \textbf{4.24} \\
mcv\_fr & 9.91 & 10.25 & 9.14 & 9.10 & 10.08 & 9.93 & \textbf{7.36} & \textbf{6.99} \\
mcv\_it & 6.86 & 7.41 & 7.04 & 7.03 & 7.78 & 7.78 & \textbf{5.39} & \textbf{4.91} \\
mcv\_pt & 7.83 & 9.27 & 8.08 & 7.54 & 9.40 & 9.57 & \textbf{5.86} & \textbf{4.91} \\
mls\_es & 10.32 & 9.62 & 11.04 & 10.10 & 12.55 & 10.32 & \textbf{7.50} & \textbf{6.30} \\
mls\_fr & 9.99 & 9.36 & 9.25 & 9.90 & 10.30 & 9.75 & \textbf{6.01} & \textbf{5.86} \\
mls\_it & 16.88 & 17.58 & 17.34 & 17.51 & 19.13 & 18.48 & \textbf{12.83} & \textbf{12.58} \\
mls\_nl & 21.51 & 22.65 & 23.01 & 23.41 & 24.32 & 23.71 & \textbf{16.44} & \textbf{15.53} \\
mls\_pl & 89.67 & 88.00 & 91.50 & 89.62 & 90.98 & 89.82 & \textbf{81.86} & \textbf{74.16} \\
mls\_pt & 26.98 & 19.87 & 27.13 & 22.80 & 34.53 & 23.97 & \textbf{16.77} & \textbf{12.75} \\
\bottomrule
\end{tabular}
\end{table*}

\begin{table*}[t]
\centering
\small
\setlength{\tabcolsep}{4pt}
\caption{Full FLEURS results across all 57 languages (CER, lower is better), with each language's \citet{joshi-etal-2020-state} resource class in parentheses. Bold marks the best score per shot setting.}
\label{tab:fleurs_full}
\begin{tabular}{l*{8}{c}}
\toprule
 & \multicolumn{2}{c}{Qwen2.5-Omni} & \multicolumn{2}{c}{+MetaSICL} & \multicolumn{2}{c}{+SFT w. same data} & \multicolumn{2}{c}{+FSA-GRPO (Ours)} \\
\cmidrule(lr){2-3}\cmidrule(lr){4-5}\cmidrule(lr){6-7}\cmidrule(lr){8-9}
Language (class) & 0-shot & 3-shot & 0-shot & 3-shot & 0-shot & 3-shot & 0-shot & 3-shot \\
\midrule
ar\_eg (5) & 16.14 & 15.53 & 18.94 & 17.14 & 19.07 & 16.16 & \textbf{12.28} & \textbf{10.62} \\
cmn\_hans\_cn (5) & 15.77 & 73.66 & \textbf{12.64} & \textbf{27.86} & 13.76 & 59.73 & 14.61 & 81.06 \\
de\_de (5) & 6.36 & 7.06 & 6.95 & 7.09 & 7.47 & 6.57 & \textbf{5.03} & \textbf{4.66} \\
en\_us (5) & 5.20 & 5.04 & \textbf{3.96} & 4.00 & 4.29 & 3.99 & 4.25 & \textbf{3.12} \\
es\_419 (5) & 5.30 & 5.63 & 6.05 & 4.84 & 7.09 & 4.67 & \textbf{5.20} & \textbf{3.57} \\
fr\_fr (5) & 7.96 & 8.58 & 7.23 & 8.31 & 8.32 & 8.12 & \textbf{6.19} & \textbf{5.98} \\
ja\_jp (5) & 17.83 & 18.11 & 18.16 & 16.83 & 17.96 & 17.61 & \textbf{15.26} & \textbf{13.47} \\
ca\_es (4) & 26.68 & 24.29 & 29.33 & 24.97 & 29.98 & 23.55 & \textbf{20.83} & \textbf{16.71} \\
cs\_cz (4) & 62.43 & 58.26 & 65.92 & 60.98 & 66.05 & 59.43 & \textbf{54.01} & \textbf{46.01} \\
fa\_ir (4) & 68.49 & 64.93 & 70.35 & 65.84 & 69.96 & 65.50 & \textbf{62.37} & \textbf{55.21} \\
fi\_fi (4) & 69.26 & 63.24 & 70.94 & 66.83 & 70.27 & 65.02 & \textbf{59.26} & \textbf{51.10} \\
hi\_in (4) & 52.55 & 51.61 & 54.04 & 52.25 & 54.42 & 52.10 & \textbf{50.69} & \textbf{48.95} \\
hr\_hr (4) & 45.34 & 39.31 & 49.21 & 41.15 & 48.60 & 39.73 & \textbf{37.20} & \textbf{28.58} \\
it\_it (4) & 5.50 & 5.61 & 5.68 & 5.21 & 5.86 & 5.75 & \textbf{4.72} & \textbf{3.93} \\
ko\_kr (4) & 11.19 & 11.59 & 11.21 & 11.58 & 11.75 & 10.74 & \textbf{9.26} & \textbf{7.89} \\
nl\_nl (4) & 10.24 & 11.43 & 10.74 & 10.07 & 10.26 & 9.43 & \textbf{7.87} & \textbf{6.30} \\
pl\_pl (4) & 59.94 & 55.80 & 61.28 & 59.20 & 63.25 & 57.38 & \textbf{51.84} & \textbf{45.10} \\
pt\_br (4) & 15.03 & 6.76 & 14.19 & 6.93 & 21.75 & 6.54 & \textbf{11.92} & \textbf{4.34} \\
ru\_ru (4) & 9.59 & 8.56 & 11.77 & 8.72 & 11.68 & 8.76 & \textbf{8.84} & \textbf{6.50} \\
sr\_rs (4) & 69.74 & 81.65 & 71.63 & 83.97 & 70.02 & 82.74 & \textbf{61.10} & \textbf{77.87} \\
tr\_tr (4) & 40.97 & 37.52 & 40.62 & 36.83 & 45.52 & 37.48 & \textbf{31.23} & \textbf{23.76} \\
vi\_vn (4) & 12.36 & 12.70 & 14.22 & 11.82 & 13.83 & 10.80 & \textbf{10.61} & \textbf{7.96} \\
be\_by (3) & 44.18 & 42.96 & 47.04 & 44.22 & 48.36 & 43.35 & \textbf{39.42} & \textbf{37.44} \\
bg\_bg (3) & 53.38 & 46.68 & 55.56 & 47.57 & 57.05 & 46.86 & \textbf{45.59} & \textbf{37.50} \\
bn\_in (3) & 75.53 & 74.08 & 77.02 & 75.89 & 76.38 & 74.43 & \textbf{71.76} & \textbf{68.55} \\
bs\_ba (3) & 49.63 & 43.77 & 53.71 & 46.05 & 53.22 & 45.45 & \textbf{42.09} & \textbf{34.27} \\
da\_dk (3) & 61.36 & 59.29 & 64.13 & 61.45 & 64.18 & 60.16 & \textbf{54.22} & \textbf{47.77} \\
et\_ee (3) & 65.36 & 62.27 & 67.13 & 67.15 & 66.76 & 64.10 & \textbf{52.04} & \textbf{48.52} \\
fil\_ph (3) & 38.12 & 36.27 & 42.99 & 37.73 & 41.25 & 36.20 & \textbf{28.18} & \textbf{24.78} \\
\bottomrule
\end{tabular}
\end{table*}

\begin{table*}[t]
\ContinuedFloat
\centering
\small
\setlength{\tabcolsep}{4pt}
\caption{Full FLEURS results across all 57 languages (continued).}
\begin{tabular}{l*{8}{c}}
\toprule
 & \multicolumn{2}{c}{Qwen2.5-Omni} & \multicolumn{2}{c}{+MetaSICL} & \multicolumn{2}{c}{+SFT w. same data} & \multicolumn{2}{c}{+FSA-GRPO (Ours)} \\
\cmidrule(lr){2-3}\cmidrule(lr){4-5}\cmidrule(lr){6-7}\cmidrule(lr){8-9}
Language (class) & 0-shot & 3-shot & 0-shot & 3-shot & 0-shot & 3-shot & 0-shot & 3-shot \\
\midrule
he\_il (3) & 84.27 & 83.25 & 80.59 & 78.13 & \textbf{79.86} & 79.33 & 81.50 & \textbf{75.84} \\
id\_id (3) & 9.50 & 9.63 & 10.53 & 10.01 & 10.89 & 9.54 & \textbf{8.80} & \textbf{6.73} \\
lt\_lt (3) & 62.45 & 58.23 & 66.94 & 61.53 & 65.82 & 60.19 & \textbf{55.23} & \textbf{47.53} \\
lv\_lv (3) & 65.09 & 60.98 & 66.61 & 63.09 & 65.86 & 62.03 & \textbf{54.96} & \textbf{50.34} \\
ro\_ro (3) & 46.68 & 43.62 & 50.97 & 46.81 & 50.78 & 47.02 & \textbf{40.14} & \textbf{33.39} \\
sk\_sk (3) & 53.43 & 44.51 & 55.85 & 45.75 & 55.39 & 44.95 & \textbf{41.12} & \textbf{32.47} \\
sl\_si (3) & 51.73 & 48.10 & 55.17 & 50.86 & 55.10 & 48.47 & \textbf{41.21} & \textbf{34.00} \\
ta\_in (3) & 80.27 & 78.10 & 83.18 & 79.50 & 83.02 & 78.36 & \textbf{80.03} & \textbf{75.35} \\
th\_th (3) & 33.15 & 32.74 & 34.57 & 34.52 & 34.26 & 33.04 & \textbf{30.84} & \textbf{29.03} \\
uk\_ua (3) & 58.02 & 37.43 & 57.01 & 37.22 & 65.01 & 37.06 & \textbf{54.62} & \textbf{28.24} \\
ur\_pk (3) & 57.48 & 55.35 & 60.41 & 56.25 & 59.97 & 54.89 & \textbf{56.31} & \textbf{47.42} \\
uz\_uz (3) & 64.33 & 54.76 & 69.29 & 56.26 & 66.55 & 54.76 & \textbf{61.36} & \textbf{44.49} \\
lo\_la (2) & 98.44 & 82.73 & 97.85 & 83.01 & \textbf{96.82} & 82.45 & 97.98 & \textbf{81.83} \\
mr\_in (2) & 68.02 & 67.58 & 69.66 & 69.16 & 69.43 & 68.40 & \textbf{64.38} & \textbf{63.13} \\
mt\_mt (2) & 55.12 & 55.43 & 59.40 & 57.23 & 57.53 & 54.95 & \textbf{47.05} & \textbf{43.88} \\
pa\_in (2) & \textbf{84.67} & 78.66 & 88.15 & 80.53 & 85.48 & 79.56 & 84.97 & \textbf{75.43} \\
sw\_ke (2) & 52.59 & 47.52 & 56.28 & 53.11 & 55.90 & 46.76 & \textbf{38.58} & \textbf{33.83} \\
as\_in (1) & 81.03 & 79.38 & 83.65 & 80.40 & 82.17 & 79.63 & \textbf{79.72} & \textbf{76.55} \\
cy\_gb (1) & 77.21 & 77.77 & 76.61 & 77.24 & 76.34 & 76.21 & \textbf{69.35} & \textbf{66.31} \\
gu\_in (1) & 81.50 & 76.14 & 83.22 & 76.90 & \textbf{80.59} & 76.42 & 81.35 & \textbf{73.59} \\
kn\_in (1) & \textbf{85.87} & 78.36 & 91.72 & 79.63 & 89.78 & 79.05 & 87.12 & \textbf{75.59} \\
mk\_mk (1) & 43.16 & 38.66 & 46.22 & 41.61 & 46.00 & 39.52 & \textbf{34.52} & \textbf{28.87} \\
ml\_in (1) & \textbf{85.05} & 81.60 & 88.64 & 82.39 & 87.96 & 82.03 & 85.52 & \textbf{79.67} \\
ne\_np (1) & 70.65 & 69.45 & 71.89 & 70.59 & 71.50 & 69.87 & \textbf{66.42} & \textbf{64.33} \\
or\_in (1) & 91.20 & 85.70 & 92.00 & 86.28 & 91.02 & 85.98 & \textbf{90.52} & \textbf{84.65} \\
sd\_in (1) & 71.77 & 65.11 & 73.27 & 67.02 & 72.27 & 65.97 & \textbf{69.17} & \textbf{62.03} \\
te\_in (1) & 80.95 & 78.90 & 84.29 & 80.53 & 81.90 & 79.32 & \textbf{79.90} & \textbf{76.40} \\
yue\_hant\_hk (1) & 41.59 & 73.34 & 40.55 & \textbf{50.45} & \textbf{39.22} & 68.43 & 40.46 & 82.77 \\
\bottomrule
\end{tabular}
\end{table*}

\end{document}